\documentclass[preprint,showpacs,showkeys,amsmath,amssymb,12pt,a4paper]{revtex4-1}
\usepackage[utf8]{inputenc}
\usepackage{graphicx}
\usepackage{amsmath}
\usepackage{hyperref}

\begin{document}

\title{Neutrinoless double beta decay and lepton flavour violation in broken $\mu-\tau$ symmetric neutrino mass models}
\author{Happy Borgohain}
\email{happy@tezu.ernet.in}
\author{Mrinal Kumar Das}
\email{mkdas@tezu.ernet.in}
\affiliation{Department of Physics, Tezpur University, Tezpur 784028, India}

\begin{abstract}

 We have studied neutrinoless double beta decay and charged lepton flavour violation in broken $\mu-\tau$ symmetric neutrino masses in a generic left-right symmetric model 
 (LRSM). The leading order $\mu-\tau$ symmetric  mass matrix originates from the type I (II) seesaw mechanism, whereas the perturbations to $\mu-\tau$ symmetry 
 in order for generation of non-zero reactor mixing angle $\theta_{13}$, as required by latest neutrino oscillation data, originates from the type II (I) seesaw
 mechanism. In our work, we considered four different realizations of $\mu-\tau$ symmetry, viz. Tribimaximal Mixing (TBM), Bimaximal Mixing (BM), Hexagonal Mixing (HM) and
 Golden Ratio Mixing (GRM). We then studied the new physics contributions to neutrinoless double beta decay (NDBD) ignoring the left-right gauge boson mixing and the heavy-light
 neutrino mixing within the framework of LRSM. We have considered the mass of the gauge bosons and 
 scalars to be around TeV and studied the effects of the new physics contributions on the effective mass and the NDBD half life and compared with the current experimental
 limit imposed by KamLAND-Zen. We further extended our analysis by correlating the lepton flavour violation of the decay processes, $\left(\mu\rightarrow 3e\right)$ and 
  $\left(\mu\rightarrow e\gamma\right)$ with the lightest neutrino mass and atmospheric mixing angle $\theta_{23}$ respectively.
\end{abstract}
\pacs{ 12.60-i, 14.60.Pq, 14.60.St}
\maketitle

\section{INTRODUCTION}

The milestone discovery of neutrino oscillation and the corresponding realization that neutrinos are massive particles has been one of the compelling revelation  which 
suggests physics beyond the Standard Model (SM). The recent neutrino experiments MINOS \cite{minos}, T2K \cite{t2k}, Double Chooz \cite{doublechooz}, Daya Bay \cite{dayabay}
and RENO \cite{reno} have not only confirmed the earlier observations but also measured the neutrino parameters more accurately. The 3$\sigma$ global fit values of the 
neutrino oscillation parameters according to recent analysis are shown in the table \ref{t1}.

\begin{table}[h!]
\centering
\begin{tabular}{||c| c| c||}
\hline
PARAMETERS & $3 \sigma$ RANGES & BEST FIT$\pm 1 \sigma$\\ \hline
$\Delta m_{21}^2[10^-5 \rm eV^2]$ & 7.11-8.18 & $7.60_{-0.18}^{+0.19}$\\ \hline
$\Delta m_{31}^2[10^-3  \rm eV^2]$(NH) & 2.30-2.65 & $2.48_{-0.07}^{+0.05}$\\
$\Delta m_{23}^2[10^-3 \rm eV^2]$(IH) & 2.26-2.48 & $2.38_{-0.06}^{+0.05}$\\ \hline
$\sin^2{\theta_{12}}$ & 0.278-0.375 & 0.323$\pm 0.016$\\ \hline
$\sin^2{\theta_{23}}$(NH) & 0.392-0.643 & ${0.567}_{-0.128}^{+0.032}$\\ 
(IH) & $0.403-0.640$ & ${0.573}_{-0.043}^{+0.025}$\\ \hline
$\sin^2{\theta_{13}}$(NH) & 0.177-0.294 & ${0.234}\pm 0.020$\\ 
(IH) & $0.183-0.297$ & {0.240}$\pm$ 0.019\\ \hline
$\delta$ & 0-2$\pi$(NH)& $254^0$\\
         & 0-2$\pi$(IH)& $266^0$\\ \hline
\end{tabular}
\caption{Global fit 3$\sigma$ values of $\nu$ oscillation parameters \cite{sigma}} \label{t1}
\end{table}

Notwithstanding, the absolute neutrino mass scale is still unperceived. However, the Planck experiment has given an upper bound on the sum of the light  neutrino mass to 
be $ \rm \sum_i\left|m_i\right|<$0.23 eV \cite{planck} in 2012 and recently the bound has been constrained to  $ \rm \sum_i\left|m_i\right|<$0.17 eV \cite{planck1}. The simplest 
hypothesis (way) to account for a neutrino mass is to introduce atleast two right handed (RH) neutrino in the Standard Model (SM). This will allow a Dirac coupling with the
Higgs, like other fermions in the SM. However, corresponding Yukawa coupling has to be fine tuned  around $ \rm 10^{-12}$ which is quite unnatural. This kind of fine
tuning can be avoided to explain the neutrino masses in the seesaw mechanism, a mechanism beyond SM (BSM) physics which is categorised into type I \cite{type1}, 
type II \cite{type2}, type III \cite{type3}, inverse \cite{inverse} seesaw mechanism. The BSM physics also unveils various phenomenon like Baryon Asymmetry of the Universe (BAU),
Lepton Number Violation (LNV), Lepton Flavour Violation (LFV), existence of dark matter etc. One of the theoretical framework to 
make the first three processes observable is the left-right symmetric model (LRSM) \cite{LRSM} which is considered to be an appealing candidate for physics BSM. Here, the gauge group 
is very simple extension of the SM gauge group. It provides a natural framework to understand the spontaneous breaking of parity and origin of small neutrino mass via
seesaw mechanism.

\par Furthermore, the physics community worldwide is embarking on the next challenging problem in finding out the nature of the neutrinos, whether they are four component
Dirac particles possesing a conserved lepton number or two component Majorana particles, along
with the absolute scale of neutrino mass. This problem is directly related to the issue of LN conservation, which is one of the most obscure sides of the SM not supported by
an underlying principle. One of such process of fundamental importance in particle physics which pops up almost in any extension of the SM is neutrinoless double beta decay
(NDBD) \cite{NDBD}. It is defined as a second order, slow radioactive process that transforms a nuclide of atomic number Z into its isobar with atomic number Z+2,
\begin{equation}\label{eq6}
\rm N\left(A,Z\right)\rightarrow N\left(A,Z+2\right)+e^-+e^-,
\end{equation}
thereby violating the total LN 
conservation. Its existence is directly linked to that of the Majorana neutrinos \cite{maj} (i.e., identical to its own anti particle).
\par The general expression for the total decay width of $ 0\nu\beta\beta$ taking into account the coulomb interaction of the electrons and the final nucleus is given by,

\begin{equation}\label{eq7}
\rm \Gamma^{0\nu}=\frac{1}{{T_{\frac{1}{2}}}^{0\nu}}=G^{0\nu}(Q,Z){\left|M^{0\nu}\right|}^2\frac{{\left|m_{\beta\beta}\right|}^2}{{m_e}^2}.
\end{equation}

The numerical
values of  $ \rm G^{0\nu}(Q,Z)$, Q and the natural abundance of several nuclei of experimental interest are given in the table \ref{t2} which are adopted from
reference \cite{table}.

\begin{table}[h!]
\centering
\begin{tabular}{||c| c| c| c||}
\hline
$\beta\beta-decay$ & $G^{0\nu}[10^{-14}y^{-1}$ & Q$[KeV]$ & Experiments\\ \hline
$48_{Ca}\rightarrow 48_{Ti}$ & 6.3 & 4273.7 & CANDLES \\ \hline
$76_{Ge}\rightarrow 76_{Se}$ & 0.63 & 2039.1 & GERDA, Majorana \\ \hline
$82_{Se}\rightarrow 82_{Kr}$ & 2.7 & 2995.5 & SuperNEMO, Lucifer \\ \hline
$100_{Mo}\rightarrow 100_{Ru}$ & 4.4 & 3035.0 & MOON, AMoRe \\ \hline
$116_{Cd}\rightarrow 116_{Sn}$ & 4.6 & 2809 & Cobra \\ \hline 
$130_{Te}\rightarrow 130_{Xe}$ & 4.1 & 2530.3 & CUORE \\ \hline
$136_{Xe}\rightarrow 136_{Ba}$ & 4.3 & 2461.9 & EXO, KamLAND-Zen, NEXT, XMASS\\ \hline
$150_{Nd}\rightarrow 150_{Sm}$ & 19.2 & 3367.3 & SNO+, DCBA/MTD \\ \hline
\end{tabular}
\caption{The values of $ \rm G^{0\nu}(Q,Z)$, Q of the initial isotope for several NDBD processes of experimental interest.} \label{t2}
\end{table}
\par The main aim of the experiment on the search for 0$ \rm \nu\beta\beta$ decay is the measurement of the effective Majorana neutrino mass, which is a combination of 
the neutrino mass eigenstates and neutrino mixing matrix terms, given by,

\begin{equation}\label{eq8}
 \rm m_{\beta\beta}=\sum_iU{_{ej}}^2m_j, \rm j=1,2,3,
\end{equation}
where, $ \rm U_{ej}$ are the elements of the first row of the neutrino mixing matrix, $ \rm U_{PMNS}$ (dependent on the known parameters $ \rm \theta_{13}, \theta_{12}$ and the unknown 
 Majorana phases $ \rm \alpha$ and $ \rm \beta$ \cite{alpha}).
$ \rm U_{PMNS}$ is the diagonalizing matrix of the light neutrino mass matrix, $m_\nu$ given by equation \ref{eq5}.

\begin{equation}\label{eq5}
\rm U_{PMNS}=\left[\begin{array}{ccc}
c_{12}c_{13}&s_{12}c_{13}&s_{13}e^{-i\delta}\\
-c_{23}s_{12}-s_{23}s_{13}c_{12}e^{i\delta}&-c_{23}c_{12}-s_{23}s_{13}s_{12}e^{i\delta}&s_{23}c_{13}\\
 s_{23}s_{12}-c_{23}s_{13}c_{12}e^{i\delta}&-s_{23}c_{12}-c_{23}s_{13}s_{12}e^{i\delta}&c_{23}c_{13}
\end{array}\right]U_{Maj}.
\end{equation}

The abbreviations used are $c_{ij}$= $\cos\theta_{ij}$, $s_{ij}$=$\sin\theta_{ij}$, $\delta$ is the Dirac CP phase while the diagonal phase matrix,
$ \rm U_{Maj}$ is $ \rm diag (1,e^{i\alpha},e^{i(\beta+\delta)}) $
\cite{P} contains the Majorana phases $ \rm \alpha$ and $ \rm \beta$. The Majorana phases $ \rm \alpha$ and $ \rm \beta$ have an effect in the process, which are allowed only if massive 
neutrinos are Majorana particles and are characterized by a violation of total LN, such as NDBD. In the standard parameterization of the mixing matrix, $ \rm m_{\beta\beta}$ is given by,
\begin{equation}\label{eq9}
\rm m_{\beta\beta}=m_1c_{12}^2c_{13}^2+m_2s_{12}^2c_{13}^2e^{2i\alpha}+m_3s_{13}^2e^{2i\beta}.
\end{equation}

A huge amount of experimental and 
theoretical activity is persued in order to detect and predict the decay process. Although no convincing experimental evidence of the decay exists till date, but new
generation of experiments that are already running or about to run assures to expedite the current limits exploring the degenerate-hierarchy region of neutrino masses.
 In addition, from the life time of this decay combined with sufficient knowledge of the nuclear matrix elements (NME), one can set a constraint involving the neutrino masses.
Moreover, if one incorporates the recent results of neutrino oscillation experiments, one can set a stringent limit on the neutrino mass scale. The latest experiments 
\cite{ndbd} that have improved the lower bound of the half life of the decay process include KamLAND-Zen \cite{kamland} and GERDA \cite{gerda} which uses Xenon-136 and 
Germanium-76 nuclei respectively. Incorporating the results from first and second phase of the experiment, KamLAND-Zen imposes the best lower limit on the decay half life using Xe-136 as
$ \rm T_{1/2}^{0\nu}>1.07\times 10^{26}$ yr at $ 90\%$ CL and the corresponding upper limit of effective Majorana mass in the range (0.061-0.165)eV.

\par Again one of the most important BSM framework to understand the origin of neutrino mass and large leptonic mixing is to identify the possible underlying symmetries. 
Symmetries can relate two or more free parameters of the model or make them vanish, making the model more predictive. The widely studied $ \rm \mu-\tau$ symmetric 
\cite{mutau}
neutrino mass matrix giving zero $ \rm \theta_{13}$ is one such scenerio where discrete flavor symmetries can relate two or more terms in the neutrino mass matrix. The neutrino 
oscillation data before the discovery of non zero $ \rm \theta_{13}$ were in perfect agreement with $ \rm \mu-\tau$ symmetric neutrino mass matrix. The four different  realizations
of neutrino mixing pattern generally found in literature which can generate from $ \rm \mu$-$\tau$ symmetric mass matrices are tribimaximal mixing (TBM), bimaximal mixing (BM), hexagonal mixing (HM), golden ratio
mixing (GRM) matrices. But, after discovery of non zero $\rm \theta_{13}$, one needs to go beyond these $ \rm \mu$-$\tau$ symmetric framework. Since the experimental value 
of $\theta_{13}$ is still much smaller than the other two mixing angles, $ \rm \mu$-$\tau$ symmetry can still be a valid approximation and the non zero $ \rm \theta_{13}$  \cite{nonzero1}
can be
accounted for by incorporating the presence of small perturbation to $ \rm \mu$-$\tau$  symmetry.
\par The discovery of neutrino oscillation has provided clear evidence of the fact that neutrinos are massive as well as the violation of the lepton flavour \cite{lfv1} 
during the
propagation of the neutrinos. Lepton flavour is consequently a broken symmetry and the SM has to be adapted to incorporate massive neutrinos and thus we
can also hope that lepton flavour violation (LFV) will be visible in the charged lepton sector \cite{lfv5}. The exact
mechanism of  LFV being unknown, its study is of large interest as it is linked to neutrino mass generation, CP violation and new physics BSM. The LFV effects from new particles 
at TeV scale are naturally generated in many models and therefore considered to be a prominent signature for new physics. In LRSM, where electroweak symmetry is broken dynamically, an 
experimentally accessible amount of LFV is predicted in a large region of parameter space. In a wide range of models for physics BSM, highest sensitivity in terms of BR is 
expected for  $ \rm \mu\rightarrow 3e $ and $ \rm \mu\rightarrow e\gamma $ decay processes.

\par To study these phenonomenon theoretically or phenomenologically, many works have been performed in LRSM based framework \cite{ls1}. In most of these works, authors
mostly considered the TBM like neutrino mass as leading order contribution and arising from type I
seesaw and using the type II seesaw as a perturbation to generate non zero $ \rm \theta_{13}$ \cite{perturbations}. More recently, the authors of \cite{ndbddb2} \cite{ndbddb1}
studied the new physics contribution to NDBD with prominent type I and type II as well as equally dominating 
type I and type II seesaw. Again, many works have been  done in charged lepton flavour violation sector in 
literature considering type I and type II dominant cases as well as equally dominant type I and type II in the  TeV scale LRSM framework which is within 
the presently accessible reach of the colliders and implements the two seesaw mechanisms naturally \cite{lfv5}.

\par In this context, we present  a phenomenological study of different $ \rm \mu-\tau$ symmetric \cite{mutau} neutrino mass  models to check their consistency with the stringent
constraints from cosmology, with various processes like LNV, LFV etc. We have taken the leading order mass matrices obeying $ \rm \mu$-$\tau$ symmetry 
 originating from type I (II) seesaw then incorporating type II (I) seesaw as perturbations to generate non zero $\theta_{13}$. Then we studied the LFV
 in the LRSM framework and further correlated the LFV of the processes $\left(\mu\rightarrow e\gamma\right)$ and $ \left(\rm \mu\rightarrow 3e \right) $
 with lightest neutrino mass and atmospheric mixing angle, $ \rm \theta_{23}$ in different neutrino mass models favouring $\mu-\tau$ symmetry. In NDBD, we discuss the different contributions \cite{ndbddb2} from  
 right handed (RH) neutrinos and RH gauge bosons, triplet Higgs \cite{higgs} as well as light heavy neutrino mixing that can contribute to the effective mass governing the
 process and identify the significant ones. In this work, we have considered only the dominant new physics contribution as coming from the diagrams containing purely RH 
 current mediated by the heavy gauge boson, $ \rm W_R$ by the exchange of heavy right handed neutrino, $ \rm N_R$ and another from the charged Higgs scalar $ \rm \Delta_R$ 
 mediated by the heavy gauge boson $ \rm W_R$ \cite{ndbddb1}. We have ignored the contributions coming from the left-right gauge boson mixing and heavy light 
 neutrino mixing.
 
\par This paper is structured as follows. In section \ref{sec:level3} we briefly discuss the left-right
symmetric model framework and the origin of neutrino mass and summarize the NDBD process in this framework in section \ref{sec:level4}. We also discuss the different 
feynmann diagrams contributing to the amplitude of the decay process (the new physics contribution) in this section. In section
 \ref{sec:level5}, we briefly discuss lepton flavor violating processes, mainly $ \rm \left(\mu\rightarrow 3e\right)$ and $ \rm \left(\mu\rightarrow e\gamma\right)$. In section \ref{sec:level6}, we present our numerical 
 analysis  and results and then in section \ref{sec:level7}, we conclude by giving a brief overview of our work.

\section{LEFT RIGHT SYMMETRIC MODEL(LRSM) AND NEUTRINO MASS}{\label{sec:level3}}

The explaination of the smallness of neutrino mass and the profile of its mixing as required by the recent experiments has been taken as a great puzzle in particle physics.
The fact that neutrino has mass implies the requirement of new physics beyond the $ \rm SU(3)_c\times SU(2)_L\times U(1)_Y$ SM
 \cite{electroweak}. One possibility to introduce neutrino mass is the so called seesaw mechanism wherein we introduce right handed heavy singlet neutrino,
$ \rm \nu_R$ (type {I} seesaw), scalar Higgs triplet (type {II} seesaw) and hypercharge-less fermion triplets (type {III} seesaw). Left-right symmetric model (LRSM) \cite{LRSM}
can be considered to be very appealing model for Physics beyond the Standard model. The seesaw mechanisms can be realized in the context of left-right symmetric model or GUTs where seesaw scale might be related to other
physical scales. 

\par In LRSM, the gauge group is a very simple extension of the standard model gauge group, $ \rm SU(3)_c\times SU(2)_L\times U(1)_Y$. Most of the problems 
like parity violation of weak interaction, masssless neutrinos, CP problems, hierarchy problems etc can be explained in the framework of LRSM, based on the gauge 
group, $ \rm SU(3)_c\times SU(2)_L\times SU(2)_R\times U(1)_{B-L}$ \cite{genericlrsm} \cite{LRSM}. In this model, the electric charge is related to the generators of the group as

\begin{equation}\label{eq14}
 \rm Q=T_{3L}+T_{3R}+\frac{B-L}{2}=T_{3L}+Y,
\end{equation}

where $ \rm T_{3L}$ and $ \rm T_{3R}$ are the 3rd components of isospin under $ \rm SU(2)_L$ and $ \rm SU(2)_R$. In LRSM, the left and right handed components of the fields are treated
on the same footing. If the Higgs sector of the model is choosen so that RH symmetry is spontaneously broken by triplets, the model gives rise to tiny neutrino masses
naturally via seesaw mechanism. Herein, there are 2 sources of lepton number violation, the Majorana masses of neutrinos and Yukawa interaction of triplet Higgs. The Quarks 
and leptons transform under the L-R symmetric gauge group as,

\begin{equation}\label{eq15}
 \rm q_L=\left[\begin{array}{c}
            u_L\\
            d_L
           \end{array}\right]\equiv \left(3,2,1,\frac{1}{3}\right),
q_R=\left[\begin{array}{c}
            u_R\\
            d_R
           \end{array}\right]\equiv \left(3,1,2,\frac{1}{3}\right) 
           \end{equation}
\begin{equation} \label{eq16}         
l_L=\left[\begin{array}{c}
            \nu_L\\
            e_L
           \end{array}\right]\equiv \left(1,2,1,-1\right),
l_R=\left[\begin{array}{c}
            \nu_R\\
            e_R
           \end{array}\right]\equiv \left(1,1,2,-1\right)                   
\end{equation}

We consider the general class of left-right symmetric model which are invariant under $ \rm SU(3)_c\times SU(2)_L\times SU(2)_R\times U(1)_{B-L}$ symmetry with the Higgs content, 
 $ \rm \phi(1,2,2,0)$, $\Delta_L(1,2,1,-1)$, $\Delta_R(1,1,2,-1)$.
A convenient representation of fields is given by 2$\times$ 2 matrices for the Higgs bidoublets and the $ \rm SU(2)_{L,R}$ triplets as,

\begin{equation}\label{eq17}
\rm \phi=\left[\begin{array}{cc}
             \phi_1^0 & \phi_1^+\\
             \phi_2^- & \phi_2^0
            \end{array}\right]\equiv \left( \phi_1,\widetilde{\phi_2}\right),   
\end{equation}

\begin{equation}\label{eq18}
 \rm \Delta_{L,R}=\left[\begin{array}{cc}
                     {\delta_\frac{L,R}{\sqrt{2}}}^+ & \delta_{L,R}^{++}\\
                     \delta_{L,R}^0 & -{\delta_\frac{L,R}{\sqrt{2}}}^+ .
                    \end{array}\right]
\end{equation}

The neutral Higgs fields $ \rm  \delta_{L,R}^0$, $ \rm \phi_1^0,\phi_2^0$ can potentially acquire VEVS $ \rm v_R,v_L,k_1,k_2$ respectively.
\begin{equation}\label{eq19}
 <\phi>=\left[\begin{array}{cc}
               \frac{k_1}{\sqrt{2}}&0\\
               0&\frac{k_2}{\sqrt{2}}
              \end{array}\right]
\end{equation}

\begin{equation}\label{eq20}
 <\Delta_{L,R}>=\left[\begin{array}{cc}
                       0 & 0\\
                       \frac{v_{L,R}}{\sqrt{2}}&0
                      \end{array}\right].
\end{equation}

The VEV $ \rm v_R$ breaks the $ \rm SU(2)_R$ symmetry and sets the mass scale for the extra gauge bosons $ \rm (W_R$ and Z$ \rm \ensuremath{'})$ and for right handed neutrino
field $ \rm (\nu_R)$. The VEVs $ \rm k_1$ and $ \rm k_2$ serves the twin purpose of breaking the remaining the  $ \rm SU(2)_L\times U(1)_{B-L}$ symmetry down to $ \rm U(1)_{em}$, thereby setting 
the mass scales for the observed $ \rm W_L$ and Z bosons and providing Dirac masses for the quarks and leptons. Clearly, $ \rm v_R$ must be significantly larger 
than $ \rm k_1$ and $ \rm k_2$ in order for $ \rm W_R$ and Z $\ensuremath{'}$ to have greater masses than the $W_L$ and Z bosons. $v_L$ is the VEV of $\Delta_L$, it plays a significant role 
in the seesaw relation which is the characteristics of the LR model and can be written as,

\begin{equation}\label{eq21}
 <\Delta_L>=v_L=\frac{\gamma k^2}{v_R}.
\end{equation}

The acceptable breaking pattern is,
$\rm SU(2)_L\times SU(2)_R\times U(1)_{B-L}\xrightarrow{<\Delta_R>} SU(2)_L\times U(1)_Y \xrightarrow{<\phi>} U(1)_{em}$.
\par The Yukawa lagrangian in the lepton sector is given by,
\begin{equation}\label{eq22}
 \rm \mathcal{L}=h_{ij}\overline{\Psi}_{L,i}\phi\Psi_{R,j}+\widetilde{h_{ij}}\overline{\Psi}_{L,i}\widetilde{\phi}\Psi_{R,j}+f_{L,ij}{\Psi_{L,i}}^TCi\sigma_2\Delta_L\Psi_{L,j}+f_{R,ij}{\Psi_{R,i}}^TCi\sigma_2\Delta_R\Psi_{R,j}+h.c.
\end{equation}

Where the family indices i, j are summed over, the indices i, j=1, 2, 3 represents the three generations of fermions. $ \rm C=i\gamma_2\gamma_0$ is the charge conjugation 
operator, $ \rm \widetilde{\phi}=\tau_2\phi^*\tau_2$ and $\gamma_{\mu}$ are the Dirac matrices. Considering discrete parity symmetry, the Majorana Yukawa couplings $f_L=f_R$ (for left-right symmetry) gives rises
to Majorana neutrino mass after electroweak symmetry breaking  when the triplet Higgs $ \rm \Delta_L$ and $ \rm \Delta_R$ acquires non zero vacuum expectation value.
Then equation (\ref{eq22}) leads to $6\times6$ neutrino mass matrix as shown in reference 2 of \cite{ls1} 

\begin{equation}\label{eq23}
\rm M_\nu=\left[\begin{array}{cc}
              M_{LL}&M_D\\
              {M_D}^T&M_{RR}
             \end{array}\right],
\end{equation}

where 
\begin{equation}\label{eq24}
\rm M_D=\frac{1}{\sqrt{2}}(k_1h+k_2\widetilde{h}), M_{LL}=\sqrt{2}v_Lf_L, M_{RR}=\sqrt{2}v_Rf_R .
\end{equation}

Where $M_D$, $M_{LL}$ and $M_{RR}$ are the Dirac neutrino mass matrix, left handed and right handed mass matrix respectively. Assuming $M_L\ll M_D\ll M_R$, the 
light neutrino mass, generated within a type I+II seesaw can be written as,

\begin{equation}\label{eq25}
 \rm M_\nu= {M_\nu}^{I}+{M_\nu}^{II},
\end{equation}

\begin{equation}\label{eq26}
\rm M_\nu=M_{LL}+M_D{M_{RR}}^{-1}{M_D}^T
      =\sqrt{2}v_Lf_L+\frac{k^2}{\sqrt{2}v_R}h_D{f_R}^{-1}{h_D}^T.
\end{equation}

Where the first and second terms in equation (\ref{eq26}) corresponds to type II seesaw and type I seesaw mediated by RH neutrino respectively.
Here,
\begin{equation}\label{eq27}
 \rm h_D=\frac{(k_1h+k_2\widetilde{h})}{\sqrt{2}k} , k=\sqrt{\left|{k_1}\right|^2+\left|{k_2}\right|^2}
\end{equation}

In the context of LRSM both type I and type II seesaw terms can be written in terms of $M_{RR}$ which arises naturally at a high energy scale as a result
of spontaneous parity breaking. In LRSM the Majorana Yukawa couplings $f_L$ and $f_R$ are same (i.e, $f_L=f_R$) and the VEV for left handed triplet $v_L$ can be written as,

\begin{equation}\label{eq28}
\rm v_L=\frac{\gamma {M_W}^2}{v_R}.
\end{equation}

Thus equation (\ref{eq26}) can be written as ,

\begin{equation}\label{eq29}
\rm M_\nu=\gamma\left(\frac{M_W}{v_R}\right)^2M_{RR}+M_D{M_{RR}}^{-1}{M_D}^T.
\end{equation}

In literature, (reference \cite{breaking} \cite{ndbddb2}) author define the dimensionless parameter $\gamma$ as,

\begin{equation}\label{eq30}
 \rm \gamma=\frac{\beta_1k_1k_2+\beta_2{k_1}^2+\beta_3{k_2}^2}{(2\rho_1-\rho_3)k^2}.
\end{equation}

Here the terms $\beta$, $\rho$ are the dimensionless parameters that appears in the expression of the Higgs potential.

\section{$0\nu\beta\beta$ decay in LRSM}{\label{sec:level4}}

Many theoretical groups has studied NDBD in connection with LRSM \cite{dbd1}. In the context  of LRSM, there are several contributions to NDBD in addition to the standard contribution via light Majorana 
neutrino exchange owing to the presence several heavy additional scaler, vector and fermionic fields . Many of the earlier works have explained it in details with the
corresponding feynmann diagrams (see ref. \cite{ndbddb2}). The various contributions to $\rm 0\nu\beta\beta$ decay 
transition rate in LRSM are briefly summarized below.
\begin{itemize}
\item Standard Model contribution to NDBD where the intermediate particles are the $\rm {W_L}^{-}$ bosons and light neutrinos. The amplitude of
this process depends upon the leptonic mixing matrix  elements elements and light neutrino masses.
\item  Heavy right handed  neutrino contribution to NDBD in which the mediator particles are the $\rm {W_L}^-$ bosons. The amplitude of 
this process depends upon the mixing between light and heavy neutrinos as well the mass of the heavy neutrino, $\rm N_i$. 
 \item Light neutrino contribution to NDBD in which the intermediate particles are $\rm {W_R}^-$ bosons. The amplitude of this process 
depends upon the mixing between light and heavy neutrinos as well as the mass of the right handed gauge boson, $\rm {W_R}^-$ boson. 
 
 \item Heavy right handed neutrino contribution  to NDBD in which the mediator paticles are the  $\rm {W_R}^-$ bosons. The amplitude of 
this process depends upon the elements of the right handed leptonic mixing matrix and the mass of the right handed gauge boson, $\rm {W_R}^-$ boson as well as the mass of the 
heavy right handed Majorana neutrino, $\rm N_i$. 
\item Light neutrino contribution from the Feynman diagram mediated by both $ \rm {W_L}^-$ and  $\rm {W_R}^-$. The amplitude of this process 
depends upon the mixing between light and heavy neutrinos, leptonic mixing matrix elements, light neutrino masses and the mass of the gauge bosons, $\rm {W_L}^-$ and $\rm {W_R}^-$.
 \item Heavy neutrino contribution from the Feynman diagram mediated by both  $ \rm {W_L}^-$ and  $\rm {W_R}^-$. The amplitude of the process 
depends upon the right handed leptonic mixing matrix elements, mixing between the light and heavy neutrinos as well as the mass of the gauge bosons, $\rm {W_L}^-$ and $\rm{W_R}^-$ 
and the mass of the heavy right handed neutrino, $\rm M_i$.
\item Triplet Higgs $\rm \Delta_L$ contribution to NDBD in which the mediator particles are $ \rm{W_L}^-$  bosons. The
amplitudes for the process  depends upon the masses of the  $ \rm{W_L}^-$ bosons, left handed triplet Higgs, $\rm \Delta_L$ as well as their coupling to leptons, $\rm f_L$.
\item Right handed triplet Higgs $\rm \Delta_R$ contribution to NDBD in which the mediator particles are $ {W_R}^-$  bosons. The amplitude
for the process depends upon the  masses of the  $ \rm{W_R}^-$ bosons, right handed triplet Higgs, $\rm \Delta_R$ as well as their  coupling to leptons, $f_R$. 
 \end{itemize}

\par However in our present work, we have considered only three of the above mentioned contributions to NDBD. One from the standard light neutrino contribution through
exchange of $\rm{W_L}^-$ as shown in figure 1(a) and the other two are the new physics contributions to NDBD which corresponds to figures 1(b) and 1(c), that is the ones mediated by ${W_R}^-$ and $\Delta_R$ respectively.
The amplitudes of the contributions are given in several earlier works like \cite{ndbddb2}.
For simple approximations, an assumption of similar mass scales for the heavy particles has been made 
in the LRSM, where, $\rm M_R\approx M_{W_R}\approx M_{\Delta_L^{++}}\approx M_{\Delta_R^{++}} \approx TeV $, at a scale accessible at the LHC. Under these assumptions, the amplitude  for the 
light-heavy mixing contribution which is proportional to $ \rm \frac{{m_D}^2}{M_R}$ remains very small (since $\rm m_\nu \approx \frac{{m_D}^2}{M_R} \approx (0.01-0.1) eV$, $m_D \approx
(10^5-10^6)$ eV which implies $ \rm \frac{m_D}{M_R} \approx (10^{-7}-10^{-6})$ eV)
Thus, we ignore the contributions involving the light and heavy
neutrino mixings. For a simplified approach, we have also ignored the mixing between $\rm{W_L}^-$ and $\rm{W_R}^-$ bosons owing to the above mentioned assumptions, which would cause a further supression 
in the amplitude of the process (for reference see \cite{ndbddb1}). 
Again, the contribution
from $\rm{\Delta_L}^-$, $\rm{W_L}^-$ is suppressed by the type II seesaw contribution to light neutrino mass and hence neglected here.

\begin{figure}
\begin{center}
\includegraphics[width=0.3\linewidth]{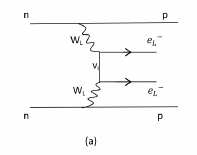}
\includegraphics[width=0.3\linewidth]{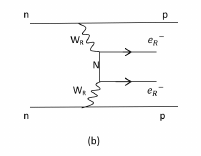}
\includegraphics[width=0.3\linewidth]{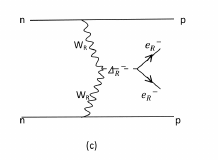}
\caption{Standard light neutrino contribution and new physics contribution ( from heavy RH neutrino and scalar Higgs triplet) to NDBD in LRSM.}
\end{center}
\end{figure}
\par Considering these contributions we have studied the NDBD. Different neutrino mass satisfying the mixing criteria namely, TBM, BM, HM and GRM are
considered as leading contribution in either type I or type II seesaw.
The perturbation is added for generation of non zero $\theta_{13}$ \cite{nonzero} in either of the seesaw terms.

 The amplitude of the corresponding processes which we have considered in our work are given by,
\begin{itemize}
\item Standard light neutrino contribution,

\begin{equation}\label{eq31}
 \rm{A_\nu}^{LL} \cong \frac{1}{{M_{W_L}}^4}\sum \frac{{U_{L_{e_i}}}^2m_i}{p^2}. 
\end{equation}

where, $\left|p\right|\sim$ 100 MeV \cite{momentum} is the typical momentum transfer at the leptonic vertex, $\rm m_{ee}=\sum{U_{L_{e_i}}}^2m_i$ 
is the effective neutrino mass. $ \rm U_{L_{e_i}}$ represents the elements of the first row of the neutrino mixing matrix, $\rm U_{PMNS}$.

\item Heavy RH neutrino contribution,

\begin{equation}\label{eq34}
\rm{A_N}^{RR} \propto \frac{1}{{M_{W_L}}^4}\frac{{U^*_{R_{e_i}}}^2}{M_i}.
\end{equation}

\item Scalar triplet contribution,

\begin{equation}\label{eq38}
\rm {A_{\Delta_R}}^{RR} \propto \frac{1}{{M_{W_R}}^4}\frac{1}{{M_{\Delta_R}}^2}f_Rv_R.
\end{equation}.
\end{itemize}
Here, $\rm \rm U^*_{R_{e_i}}$ denotes the first row of the unitary matrix diagonalizing the right handed neutrino mass matrix, $ \rm M_{RR}$ with mass eigen values, $ \rm M_i$. 

\section{Lepton Flavour Violation (LFV)}{\label{sec:level5}}
There have been various attempts to observe and predict
theoretically the manifestation of LFV involving various modes of muon decay since long. The most promising LFV low energy channels are probably $\mu\rightarrow e\gamma$, 
$\mu\rightarrow 3e$, $\mu\rightarrow e$ conversion in nuclei which occur in rates accessible in recent experiments. Defining the decay rates (from reference \cite{lfv1})as,

\begin{equation}\label{eq39}
 \rm\Gamma_{\mu}\equiv \Gamma\left(\mu^{-}\rightarrow e^{-}\nu_{\mu}\overline{\nu_{e}}\right), \Gamma_{capt}^{Z}\equiv \Gamma\left(\mu^{-}+A\left(Z,N\right)\rightarrow \nu_{\mu}+A\left(Z-1,N+1\right)\right).
\end{equation}

\par The relevant branching ratios (BR) for the processes are,

\begin{equation}\label{eq40}
 \rm BR_{\mu\rightarrow e\gamma}\equiv \frac{\Gamma\left(\mu^{+}\rightarrow e^{+}\gamma\right)}{\Gamma_{\mu}},
\end{equation}

\begin{equation}\label{eq41}
\rm  BR_{\mu\rightarrow e}^{Z}\equiv \frac{\Gamma\left(\mu^{-}+A\left(N,Z\right)\rightarrow e^{-}+A\left(N,Z\right)\right)}{\Gamma_{capt}^{Z}},
\end{equation}

\begin{equation}\label{eq42}
 \rm BR_{\mu\rightarrow 3e}\equiv \frac{\Gamma\left(\mu^{+}\rightarrow e^{+} e^{-} e^{+}\right)}{\Gamma_{\nu}}.
\end{equation}

The selected limits for lepton flavour violating muon decays and muon to electron conversion experiments are shown in table \ref{t3}

 \begin{table}[h!]
\centering
\begin{tabular}{||c| c| c||}
\hline
DECAY CHANNEL & EXPERIMENT & BRANCHING RATIO LIMIT \\ \hline
$\mu\rightarrow e\gamma$  & MEG & $ <4.2\times 10^{-13}$  \cite{muegamma}\\
$\mu\rightarrow $eee  & SINDRUM & $<1.0\times 10^{-12}$  \cite{SINDRUM}\\
$\mu Au\rightarrow$ e Au & SINDRUM II & $<7\times 10^{-13}$  \cite{sindrum2}\\ \hline
\end{tabular}
\caption{Experimental limits on LFV muon decays.}\label{t3}
\end{table}

\par In the SM seesaw, the LFV decay rates induced by neutrino mixing are suppressed by tiny neutrino masses,$\left(\frac{{\Delta m_A}^2}{{M_W}^2}\right)\sim10^{-50}$
and hence are well below the current experimental limits and even the distant future sensitivities. New physics beyond the standard model is required to make the process observable,
there are several theoretical frameworks BSM that could provide the necessary operators. One of those theories is the LRSM in which several new
contributions appear due to the additional RH current interactions, which could lead to sizeable LFV rates for TeV scale $v_R$ that occur at rates observable in current experiments.
LFV in the LRSM has been studied in many previous works. There are various LFV processes providing constraints on the masses of the right handed neutrinos and doubly charged scalars. It turns out that  the process
$\mu\rightarrow 3e $ induced by doubly charged bosons $\Delta_L^{++}$ and $\Delta_R^{++}$  and $\mu\rightarrow e\gamma$ provides the most relevant constraint.  
In our present work, we consider these processes in minimal left-right symmetric model (MLRSM). The limit of branching 
ratio of the process  $\mu\rightarrow 3e $ as shown in table \ref{t3} is $<1.0\times 10^{-12}$ at $ 90\%$ CL was obtained at the Paul Scherrer institute (PSI) over 20 years ago by the SINDRUM 
experiment \cite{SINDRUM}. Presently the Mu3e collaboration has submitted a letter of intent to PSI to perform a new  improved search for the decay $\mu\rightarrow 3e $ with
a sensitivity of $10^{-16}$  at $95\%$ CL \cite{sindrum2} which corresponds to an improvement by four orders of magnitude compared to the former SINDRUM experiment. Whereas the 
new upper limit for BR of the process $\mu\rightarrow e\gamma$ is established to be $<4.2\times 10^{-13}$ at $ 90\%$ CL by the MEG collaboration. Taking 
into account the contributions from heavy righthanded neutrinos and Higgs scalars, the expected branching ratios and conversion rates of the above processes have been 
calculated in the LRSM in the work (first reference in \cite{lfv}). 
\par The BR for the
process $(\mu\rightarrow 3e )$ is given by, \par
\begin{equation}\label{eq43}
 \rm BR\left(\mu\rightarrow 3e\right)=\frac{1}{2}{\left|h_{\mu e}h_{ee}^{*}\right|}^{2}\left(\frac{{m_{W_L}}^4}{{M_{\Delta_L}^{++}}^4}+\frac{{m_{W_R}}^4}{{M_{\Delta_R}^{++}}^4}\right).
\end{equation}
 Where $h_{ij}$ describes the lepton Higgs coupling in LRSM and is given by,

 \begin{equation}\label{eq44}
  \rm h_{ij}=\sum_{n=1}^{3}V_{in}V_{jn}\left(\frac{M_n}{M_{W_R}}\right), i,  j=e,\mu,\tau.
 \end{equation}
  \par For $\mu\rightarrow e\gamma$, the relevant BR is given by, \cite{lfv1}
  \begin{equation}\label{eqa}
   \rm BR\left(\mu\rightarrow e\gamma\right)= 1.5\times 10^{-7}{\left|g_{lfv}\right|}^2{\left(\frac{1 TeV}{M_{W_R}}\right)}^4,
   \end{equation}\\
  where, $ g_{lfv}$ is defined as,
  \begin{equation}\label{eqb}
   \rm g_{lfv}=\sum_{n=1}^{3}V_{\mu n}{V_{e n}}^*{\left(\frac{M_n}{M_{W_R}}\right)}^2=\frac{\left[M_R {M_R}^*\right]_{\mu e}}{{M_{W_R}}^2}
  \end{equation}

\par The sum is over the heavy neutrinos only. $ \rm M_{\Delta_{L,R}}^{++}$ are the masses of the doubly charged bosons, $ \rm {\Delta_{L,R}}^{++}$, V is the mixing matrix
 of the right handed neutrinos with the electrons and muons. $ \rm M_n(n=1,2,3) $ are the right handed neutrino masses.

\section{NUMERICAL ANALYSIS AND RESULTS}{\label{sec:level6}}
In our present work we have studied LNV (NDBD) for standard as well as non standard contributions for the effective mass as well as the half life governing the decay process in 
the framework of LRSM. We have also correlated the LFV of the process, $\rm \mu\rightarrow 3e $ and $\rm \mu\rightarrow e\gamma $   with the lightest neutrino mass and 
atmospheric mixing angle, $\rm \theta_{23}$ respectively for both normal and inverted mass hierarchies. In this section we present a detailed analysis of our work and we 
have divided it into different subsections, firstly the standard light 
neutrino contribution to NDBD and then the new physics contribution to NDBD considering perturbation in type II and then type I seesaw. Lastly we have shown the analysis of 
correlating LFV with $\rm m_{lightest}$ and $\rm \theta_{23}$.

\subsection{Standard light neutrino contribution}
For NDBD mediated by the light Majorana neutrinos, the half life of the decay process is given by equation (\ref{eq6}) and the effective mass governing the process
is as given in equation (\ref{eq9}).
 In our present work, we first evaluated the effective light neutrino mass within the standard mechanism using the formula (\ref{eq8})
 where, $\rm U_{ej}$ are the elements of the first row of the neutrino mixing matrix, $ \rm U_{PMNS}$ (dependent on the known parameters $ \rm \theta_{13}, \theta_{12}$ and the unknown 
 Majorana phases $\rm \alpha$ and $\rm \beta$).
$ \rm U_{PMNS}$ is the diagonalizing matrix of the light neutrino mass matrix, $ \rm m_\nu$, such that 
\begin{equation}\label{eq47}
 \rm m_\nu= U_{PMNS}{M_\nu}^{(diag)} {U_{PMNS}}^T,
\end{equation}
where $ \rm {M_\nu}^{(diag)}=diag(m_1,m_2,m_3)$.
In the case of 3 neutrino mixing, 2 $\nu$ mass spectra are possible,

\begin{itemize}
\item Normal Hierachy $ \rm (NH)$ which corresponds to $ m_1< m_2\ll m_3$ ; $\Delta {m_{12}}^2\ll\Delta {m_{23}}^2$.
\item Inverted Hierarchy $ \rm (IH)$ which corresponds to $m_3\ll m_1\sim m_2$ ; $\Delta {m_{12}}^2\ll\left|\Delta {m_{13}}^2\right|$.
\end{itemize}
\par In both the spectra, $ \rm \Delta {m_{12}}^2=\Delta {m_{solar}}^2$. For NH, $ \rm \Delta {m_{23}}^2=\Delta {m_{atm}}^2$ and for IH, $\left|\Delta {m_{13}}^2\right|=\Delta {m_{atm}}^2$.
In the case of NH, the neutrino masses $m_2$ and $m_3$ are connected with the lightest mass $m_1$ by the relation,

\begin{equation}\label{eq10} 
 \rm m_2=\sqrt{m_1^2+\Delta m_{sol}^2},m_3=\sqrt{m_1^2+\Delta m_{sol}^2+\Delta m_{atm}^2}.
\end{equation}

In IH, $ \rm m_3$ is the lightest mass and we have,

\begin{equation}\label{eq11} 
\rm m_1=\sqrt{m_3^2+\Delta m_{atm}^2},m_2=\sqrt{m_3^2+\Delta m_{sol}^2+\Delta m_{atm}^2}.  
\end{equation}

For both the normal and inverted hierarchies, equation (\ref{eq9})
can be written in terms of lightest neutrino mass as,

for NH,
\begin{equation}\label{eq12} 
\rm m_{\beta\beta}=m_1{c_{12}}^2{c_{13}}^2+\sqrt{(m_1^2+\Delta m_{sol}^2}{s_{12}}^2{c_{13}}^2e^{2i\alpha})+
 \sqrt{(m_1^2+\Delta m_{sol}^2+\Delta m_{atm}^2}{s_{13}}^2e^{2i\beta}),
\end{equation}

for IH,
\begin{equation}\label{eq13} 
 \rm m_{\beta\beta}=\sqrt({m_3^2+\Delta m_{atm}^2}{c_{12}}^2{c_{13}}^2)+\sqrt{(m_3^2+\Delta m_{sol}^2+\Delta m_{atm}^2}{s_{12}}^2{c_{13}}^2e^{2i\alpha}+m_3{s_{13}}^2e^{2i\beta}).
\end{equation}

\par
\begin{figure}[h!]
\includegraphics[width=0.49\textwidth]{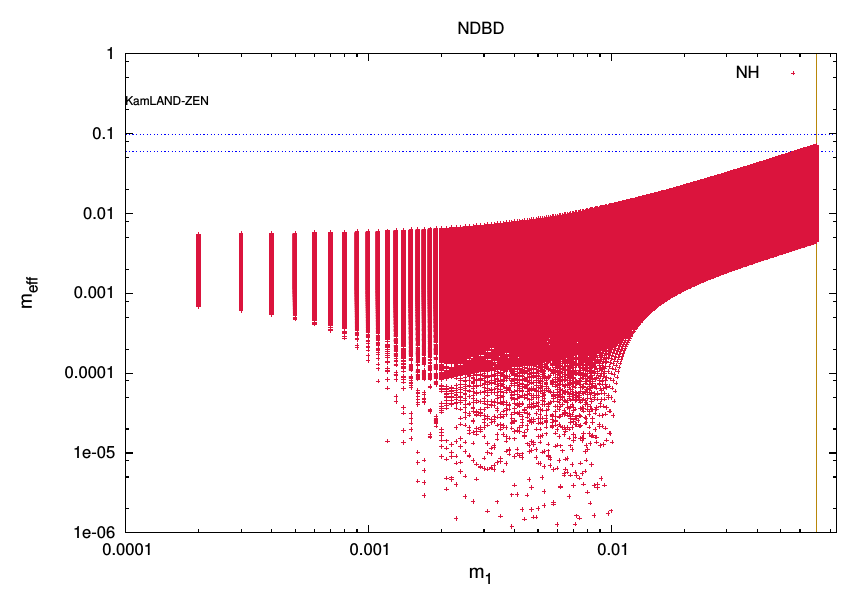}
\includegraphics[width=0.49\textwidth]{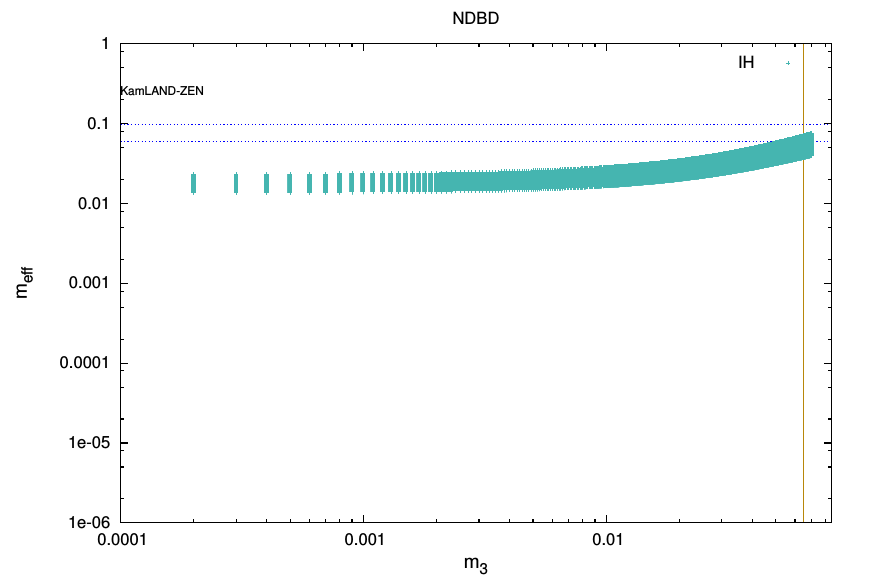}
\caption{Effective Majorana mass for 0$\nu\beta\beta$ as a function of lightest neutrino mass, $\rm m_1$ (in eV) for NH (as shown in figure left) and $\rm m_3$ (in eV)for IH (as shown in figure 
right) within the standard mechanism. The blue dashed line and the yellow solid line represents the KamLAND-Zen bound on the effective mass and the Planck bound on the sum of 
the absolute neutrino mass respectively.} \label{fig6}
\end{figure}

 The $3\sigma$ ranges of the mass squared 
differences and mixing angles from global analysis of oscillation data are outlined as in the table \ref{t1}.
 Using the best fit values of the mass squared differences and the $\rm 3\sigma$ ranges of the three mixing angles from a global analysis of oscillation data (as shown 
in table \ref{t1}), we have shown the variation of the effective Majorana mass as a function of the lightest neutrino mass $ \rm m_1$ (for NH) and $ \rm m_3$ (for IH). During our 
calculation, we have varied the Majorana phase $\rm \alpha$ and $\rm \beta$ from 0 to $\rm 2\pi$.
The effective mass assumes different values depending on whether the neutrino mass states follows normal hierarchy (NH) or inverted hierarchy (IH). We have used equations 
(\ref{eq12})and (\ref{eq13}) in evaluating the effective mass in terms of the lightest neurino mass. 
The variation is shown in figure \ref{fig6}. It is seen from the figure that the light neutrino contribution to neutrinoless double beta decay $(0\nu\beta\beta)$ can 
saturate the bound  imposed by KamLAND-ZEN $ \rm (\leq 0.061-0.165  eV)$ [reference (\cite{kamland})] only for the higher values of lightest neutrino masses which is disallowed by the Planck 
data (lightest mass for NH $\sim$ 0.07 and lightest mass for IH $\sim$ 0.065). 

\par Again, we have evaluated the effective majorana mass for different leptonic
mixing patterns possessing $\rm \mu-\tau$ symmetry, namely, tribimaximal, golden ratio and hexagonal mixing \cite{mutau}using equation (\ref{eq12})and (\ref{eq13}).
In all the different $\rm \mu-\tau$ symmetric mixing patterns which we have considered, i.e., TBM, HM, GRM, the reactor mixing angle $\theta_{13}$ is 0 and $\theta_{23}$ is $45^0$.
Whereas  $\rm \theta_{12}$=$35.5^0$,(for TBM), $\theta_{12}=30^0$(for HM), $\theta_{12}$=$31.71^0$(for GRM). Since,
$\rm \theta_{12}$=$45^0$, i.e, BM has been ruled out by experiments, we have ignored this case for the standard contributions. Again, it
is to be noted that there are two values of $\rm\theta_{12}$ for GRM, which are, $31.7^0$ and $35.96^0$ \cite{GRM}. In our present study, we have considered the first value
which is allowed as mentioned in reference \cite{GRM}\cite{GRM1}

\par
\begin{figure}[h!]
\includegraphics[width=0.82\textwidth]{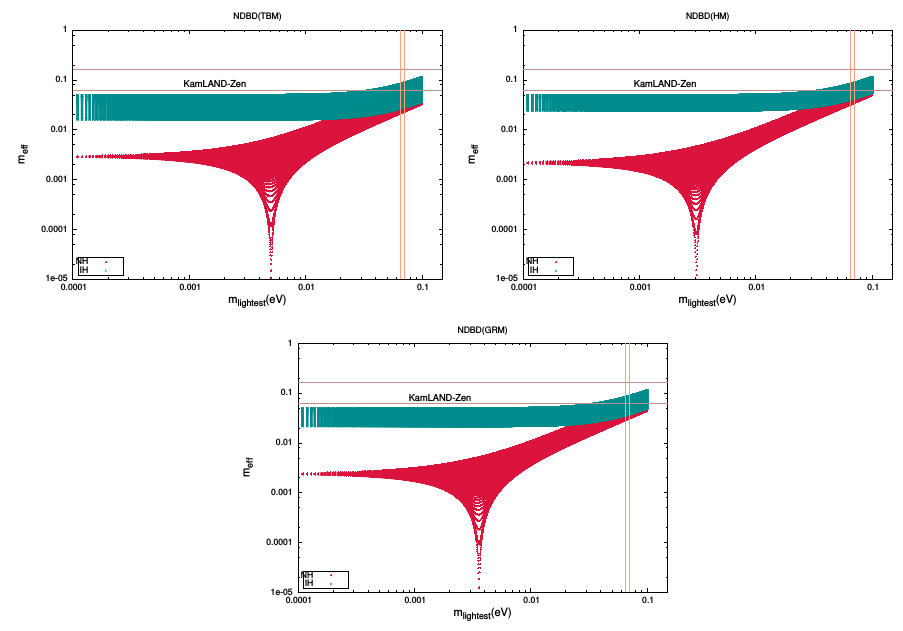}
\caption{Standard light neutrino  contribution to effective mass for 0$\nu\beta\beta$ for different neutrino mass models (TBM, HM and GRM) as a function of lightest neutrino mass (in eV) for NH/IH (m1/m3)
The horizontal lines represents the upper limit of effective mass propounded by kamLAND-Zen and vertical line represents the plancks bound on lightest neutrino mass for NH and IH.}\label{fig7}.
\end{figure}

 The variations of $ \rm m_{\nu}^{eff}$ for the different mixing patterns for NH and IH in terms of lightest neutrino mass are shown as in figure \ref{fig7}.

\subsection{New physics contribution to NDBD  considering perturbation in type II seesaw.}
For the new Physics contribution, we have considered the contributions of $ \rm 0\nu\beta\beta$ from the right handed current and from the triplet Higgs $ \rm (\Delta_R)$. The
contributions from the left handed Higgs triplet, $ \rm \Delta_L$ is suppressed by the light neutrino mass. Also we consider the mixing between LH and RH sector to be so small 
that their contributions to $0\nu\beta\beta$ can be neglected. The total effective mass is thus given by the formula, (as in \cite{ndbddb1})
\begin{equation}\label{eq55}
 \rm {m_{N+\Delta_R}}^{eff}=p^2\frac{{M_{W_L}}^4}{{M_{W_R}}^4}\frac{{U_{Rei}}^*2}{M_i}+p^2\frac{{M_{W_L}}^4}{{M_{W_R}}^4}\frac{{{U_{Rei}}^2}M_i}{{M_{\Delta_R}}^2}.
\end{equation}
\par Here, $ \rm <p^2> = m_e m_p \frac{M_N}{M_\nu}$ is the typical momentum exchange of the process, where $ \rm m_p$ and $ \rm m_e$ are the mass of the proton and electron respectively 
and $ \rm M_N$ is the NME corresponding to the RH neutrino exchange.
We know that TeV scale LRSM plays an important role in 0$\nu\beta\beta$ decay. We have considered the values $ \rm M_{W_R}$ = 3.5 TeV, $ \rm M_{W_L}$ = 80 GeV, $ \rm M_{\Delta_R}\approx $3TeV,
the heavy RH neutrino $\approx$ TeV which are within the recent collider limits \cite{collider}. 
The allowed value of p (the virtuality of the exchanged neutrino) is in the range
$\sim $ (100-200) MeV. In our analysis, we have taken p$\simeq$180 MeV \cite{ndbddb2}.

\par Thus,
\begin{equation}\label{eq56}
 \rm p^2\frac{{M_{W_L}}^4}{{M_{W_R}}^4} \simeq 10^{10} {eV}^2.
\end{equation}
However, equation (\ref{eq55}) is valid only in the limit $ \rm {M_i}^2 \gg\left|<p^2>\right|$ and $ \rm {M_\Delta}^2\gg\left|<p^2>\right|$. 
\par  The formula for light $\nu$ masses in
the presence of both type I and type II seesaw can be written as,
\begin{equation}\label{eq57}
 \rm M_\nu={ M_\nu}^{I}+{ M_\nu}^{II},
\end{equation}
\begin{equation}\label{eq58}
 \rm U_{PMNS}{M_\nu}^{(diag)} {U_{PMNS}}^T={ M_\nu}^{II}+U_{(\mu-\tau)}U_{Maj}{M_\nu}^{I(diag)}{U_{Maj}}^T{U_{(\mu-\tau)}}^T,
\end{equation}
where, $ \rm U_{PMNS}$ and $U_{(\mu-\tau)}$ represents the diagonalizing matrix of $ \rm M_\nu$ and $ \rm {M_\nu}^{I}$. The Majorana phases have been taken in the type
I seesaw term \cite{ndbddb1}. From equation(\ref{eq58}) we can evaluate $ \rm { M_\nu}^{II}$. We have considered the case when $ \rm { M_\nu}^{I}$ possess
$ \rm \mu-\tau$ symmetry, with the various choices 
for mixing matrices such as TBM, BM, HM, GRM, with uniquely predicting $ \rm \theta_{13}=0$. We have considered $ \rm {M_\nu}^{I(diag)}=X{M_\nu}^{(diag)}$, where 
we have introduced the parameter X to describe the reltive strength of the type I and II seesaw terms. The parameter X can take any numerical value
provided the two seesaw terms gives rise to correct light neutrino mass matrix. In our case, we have considered X=0.5, i.e., equal contributions from both the seesaw terms.
The required correction to $\mu-\tau$ type $\nu$ mass matrix for generation of non
zero reactor mixing angle ($ \rm \theta_{13}$) can be obtained from the perturbation matrix, $ \rm {M_\nu}^{II}$ mass matrix. $ \rm {M_\nu}^{II}$ can 
be constructed as,
 \begin{equation}
\rm {M_\nu}^{II}=\left[\begin{array}{ccc}
S_{11}&S_{12}&S_{13}\\
S_{21}&S_{22}&S_{23}\\
S_{31}&S_{32}&S_{33}
\end{array}\right].
\end{equation}
It can be derived using equation (\ref{eq57}). The type II seesaw mass matrix is evaluated in terms of light neutrino mass matrix, constructed using the best fit neutrino data
and $\mu-\tau$ symmetric type I mass matrices (TBM, BM, HM, GRM). The elements are shown in appendix.

 To evaluate $ \rm {m_{N+\Delta_R}}^{eff}$, we need the
diagonalizing matrix of the heavy right handed Majorana mass matrix $ \rm M_{RR}$, $U_{Rei}$ and its mass eigenvalues, $ \rm M_i$.
\par $ \rm M_{RR}$ can be written in the form(from reference \cite{newphysics}) and is evident from equation (\ref{eq29})
\begin{equation}\label{eq59}
 \rm M_{RR}=\frac{1}{\gamma}{\left(\frac{v_R}{M_{W_L}}\right)}^2{ M_\nu}^{II},
\end{equation}
\begin{equation}\label{a}
\rm  { M_\nu}^{II}= U_{PMNS}{M_\nu}^{(diag)} {U_{PMNS}}^T-U_{(\mu-\tau)}U_{Maj}{M_\nu}^{I(diag)}{U_{Maj}}^T{U_{(\mu-\tau)}}^T.
\end{equation}
\par
\begin{figure}[h!]
\includegraphics[width=0.7\textwidth]{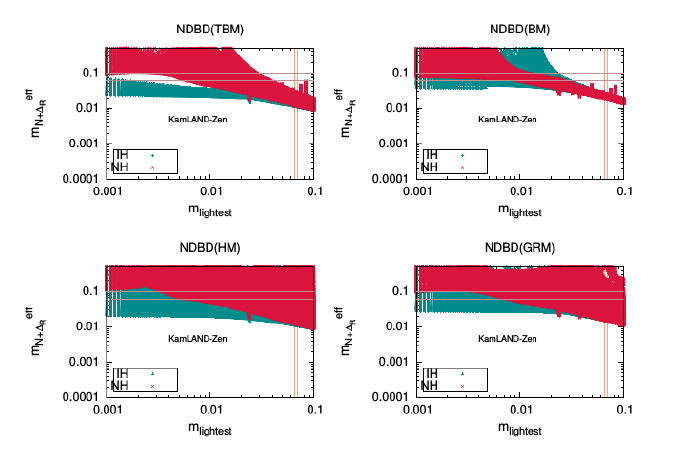}
\caption{New Physics contribution to effective mass for 0$\nu\beta\beta$  considering perturbation in type II seesaw for different mass models (TBM, BM, HM and GRM).}\label{fig8}
\end{figure}

 In the above equation, $ \rm U_{(\mu-\tau)}$ represents $ \rm U_{TBM}$, $ \rm U_{BM}$, $ \rm U_{HM}$, $ \rm U_{GRM}$ \cite{mutau}, i.e, the diagonalizing matrices of the TBM, BM, HM and GRM mass
matrices. 
For TeV scale type I + type II seesaw, we have fine tuned the dimensionless parameter, $\gamma\sim 10^{-10}$, we have considered $v_R\sim TeV$. Thus 
after obtaining $ \rm M_{RR}$, we diagonalized it and obtained the eigenvalues, $ \rm M_i$ and its diagonalizing matrix in terms of the lightest neutrino 
mass ( $ \rm m_1$  or $ \rm m_3$)for (NH/IH) and the Majorana phases ($ \rm \alpha$ and $\beta$). We have varied the Majorana phases $ \rm \alpha$ and $ \rm \beta$ from 0 to $ 2\pi$
and evaluated the effective mass for new physics contribution using formula (\ref{eq55}) in terms of lightest neutrino mass. This is shown in figure (\ref{fig8}).We 
have imposed the KamLAND-Zen bound on the new physics contribution to effective mass and the Planck bound on the sum of the absolute neutrino mass.

\subsection{New physics contribution to NDBD  considering perturbation in type I seesaw.}
 Alternatively, we have again considered the type {II} seesaw to give rise to $\rm \mu-\tau$ type neutrino
mass matrix and the necessary correction to obtain non-zero $\rm \theta_{13}$ is obtained from the type {I} seesaw term. Thus, $\rm { M_\nu}^{II}$ in equation (\ref{eq59})
can be written as,
\begin{equation}\label{eq60}
\rm { M_\nu}^{II}=U_{(\mu-\tau)}U_{Maj}{M_\nu}^{{II}(diag)}{U_{Maj}}^T{U_{(\mu-\tau)}}^T,
\end{equation}
\par where, $\rm U_{(\mu-\tau)}$ represents $\rm U_{TBM},U_{BM},U_{HM},U_{GRM}$.
\begin{equation}\label{eq61}
 \rm {M_\nu}^{I}= M_\nu-{ M_\nu}^{II},
\end{equation}
\begin{equation}\label{eq62}
  \rm{M_\nu}^{I}= U_{PMNS}{M_\nu}^{(diag)} {U_{PMNS}}^T-{ M_\nu}^{II}.
\end{equation}

\begin{figure}[h!]
\includegraphics[width=0.7\textwidth]{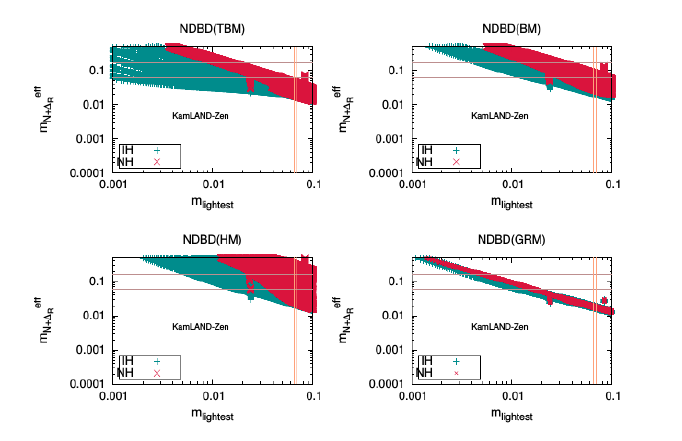}
\caption{New Physics contribution to effective mass for 0$\nu\beta\beta$ considering perturbation in type I seesaw for different mass models (TBM, BM, HM and GRM).}\label{fig9}
\end{figure}

Like in the previous case, we have again evaluated the right handed Majorana mass matrix using equation (\ref{eq59}). We  have fine tuned the dimensionless parameter $\gamma$ 
and then by diagonalizing the right handed Majorana mass matrix $ \rm M_{RR}$, we have obtained $ \rm U_{Rei}$ and the eigenvalues, $ \rm M_i$ $\rm (i.e. {M_{RR}}^{(diag)})$
where,
\begin{equation}\label{eq63}
 \rm M_{RR}= U_{Rei}{M_{RR}}^{(diag)} {U_{Rei}}^T 
\end{equation}
We then evaluated the effective Majorana mass, $ \rm {m_{N+\Delta_R}}^{eff}$ using equation (\ref{eq55}) as a function of the lightest left handed neutrino mass.
This is shown in figure (\ref{fig9}). When we consider the type II seesaw term to be $ \rm \mu-\tau$ symmetric and the perturbation from the type I seesaw term, the type I seesaw
mass matrix can be derived as in the previous case  and is shown in appendix.
\par
\begin{figure}[h!]
\includegraphics[width=0.64\textwidth]{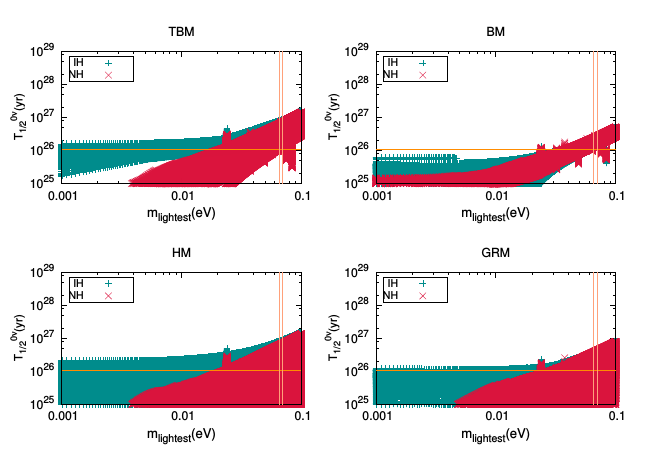}
\caption{New Physics contributions to half life of 0$\nu\beta\beta$  considering perturbation in type II seesaw in different mass 
models (TBM, BM, GRM, HM) for normal and inverted hierarchies. The horizontal line represents the lower limit on 0$\nu\beta\beta$ half life 
imposed by KamLAND-ZEN projected sensitivity respectively.}\label{fig10}
\end{figure}

\begin{figure}[h!]
\includegraphics[width=0.64\textwidth]{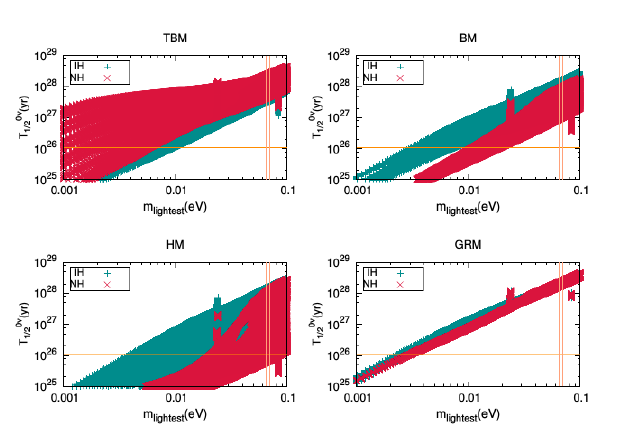}
\caption{New Physics contributions to half life of 0$\nu\beta\beta$ considering perturbation in type I seesaw in different mass 
models (TBM, BM, GRM, HM) for normal and inverted hierarchies. The horizontal line represents the lower limit on 0$\nu\beta\beta$ half life 
imposed by KamLAND-ZEN projected sensitivity respectively.}\label{fig11}
\end{figure}

 For the new physics contribution in which the type II term acts as the perturbation, we have also evaluated
the half life of the $0\nu\beta\beta$ decay process using equation (\ref{eq7}), where
\begin{equation}\label{eq64}
\rm {\left| {m_\nu}^{eff}\right|}^2={\left|{m_N}^{eff}+{m_{\Delta_R}}^{eff}\right|}^2.
\end{equation}
By substituting the values of the phase factors(${G_0}^\nu$) \cite{phasefactor} \cite{ndbdg}, nuclear matrix element(NME) \cite{nme} \cite{ndbdg} and mass
of electron, we have obtained the half life as a function of the lightest mass in the different mixing patterns for both NH and IH, as shown in figure (\ref{fig10}). In the similar process, we have also
computed the half life  for new physics contribution to NDBD in which the type I term acts as the perturbation, for generation of non zero $\theta_{13}$.
It is shown in figure (\ref{fig11}).

\subsection{Correlating LFV with lightest neutrino mass and $ \rm \theta_{23}$ }
\begin{figure}[h!]
\includegraphics[width=0.6\textwidth]{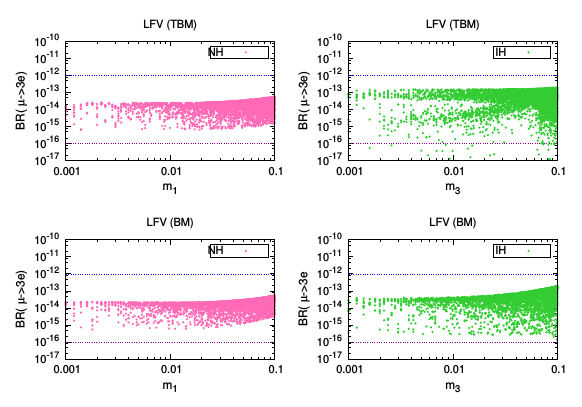}
\caption{Total contribution to lepton flavour violation  shown as a function of the lightest neutrino mass for the 
TBM and BM neutrino mass models for normal and inverted hierarchies. The blue and violet dashed line shows the limit of BR as given by SINDRUM 
experiment and the recently proposed limit of $\mu$ 3e experiment respectively.}
\label{fig12}
\end{figure}

\begin{figure}[h!]
\includegraphics[width=0.6\textwidth]{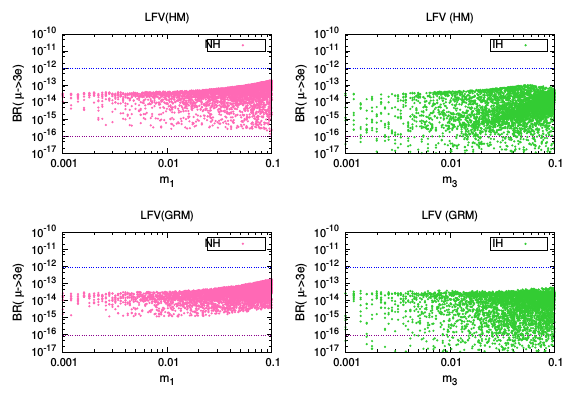}
\caption{Total contribution to lepton flavour violation  with type (I+II) seesaw shown as a function of the lightest neutrino mass for the HM
and GRM neutrino mass models for normal and inverted hierarchies. The blue and violet dashed line shows the limit of BR as given by SINDRUM 
experiment and the recently proposed limit of $\mu$ 3e experiment respectively.}\label{fig13}
\end{figure}

\begin{figure}[h!]
\includegraphics[width=0.6\textwidth]{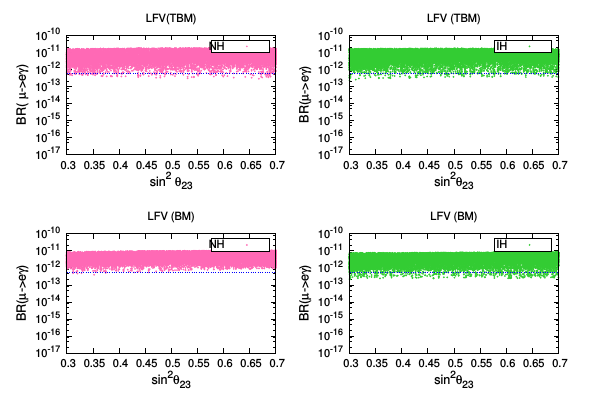}
\caption{Total contribution to lepton flavour violation  with type (I+II) seesaw shown as a function of the atmospheric mixing angle $\theta_{23}$ for TBM
and BM neutrino mass models for normal and inverted hierarchies. The blue  dashed line shows the limit of BR.}\label{fig14}
\end{figure}

\begin{figure}[h!]
\includegraphics[width=0.6\textwidth]{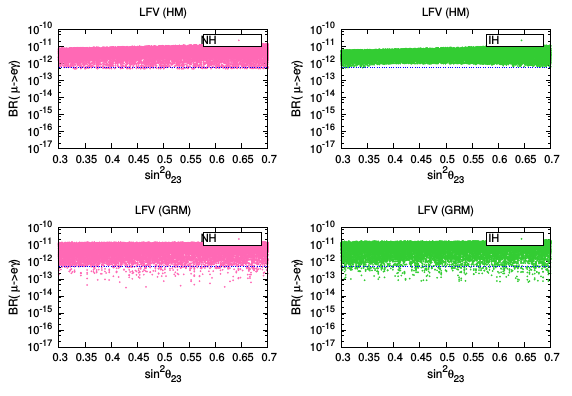}
\caption{Total contribution to lepton flavour violation  with type (I+II) seesaw shown as a function of the atmospheric mixing angle $\theta_{23}$ for the HM 
and GRM neutrino mass models for normal and inverted hierarchies. The blue  dashed line shows the limit of BR.}\label{fig15}
\end{figure}

 To correlate LFV with neutrino mass in our analysis, we have considered the LFV processes, $\rm \mu\rightarrow 3e $ and $\rm \mu\rightarrow e\gamma$. The BR for both the processes have a strong flavour dependence
 on the RH mixing matrix. Since the process $\rm \mu\rightarrow 3e $ is controlled by $\rm h_{\mu e}h_{ee}^{*}$ whereas $\rm \mu\rightarrow e\gamma$ is controlled by the factor
$\rm \left[M_R {M_R}^*\right]_{\mu e}$ , the later is independent of the Majorana CP phases and the lightest neutrino mass, $\rm m_{j}$. We have correlated the BR of the process
$\rm \mu\rightarrow 3e $  with the lightest neutrino mass $\rm(m_1/m_3)$ for (NH/IH). The BR of the process $\rm \mu\rightarrow e\gamma$ is correlated with the atmospheric mixing angle, $\rm\theta_{23}$, since the other
two mixing angles $\rm \theta_{12}$ and $\rm \theta_{13}$ are measured precisely.
For calculating the BR, we used the expression 
given in equation (\ref{eq43}) and (\ref{eqa}). The lepton Higgs coupling $\rm h_{ij}$ in (\ref{eq44}) can be computed explicitly for a given RH neutrino mass matrix as shown in
equation (\ref{eq59}) by diagonalizing the RH neutrino mass matrix and obtaining the mixing matrix element, $V_i$ and the eigenvalues $M_i$. For evaluating $M_{RR}$, we 
need to know $\rm{ M_\nu}^{II} $, as evident from equation (\ref{eq59}). We computed ${ M_\nu}^{II} $ from equation (\ref{eq58}). For determining the BR for $\mu\rightarrow 3e$, 
we imposed the best fit values of
the parameters, $\rm{\Delta m_{sol}}^2$, $\rm{\Delta m_{atm}}^2$, $\rm\delta$, $\rm\theta_{13}$, $\rm\theta_{23}$, $\rm\theta_{12}$ in $\rm M_\nu$ . The numerical values of $\rm{ M_\nu}^{I} $ can be computed 
as before for different mixing patterns, TBM, BM, HM, GRM. Thus, we get $\rm{ M_\nu}^{II} $ as a function of the parameters $\rm\alpha,\beta$ and $\rm m_{lightest}$. 
Then varying both the Majorana phases, $\rm\alpha, \beta$ from 0 to 2$\pi$, we obtained $\rm{ M_\nu}^{II} $ as a function of $m_{lightest}$. Similarly, for $\mu\rightarrow e\gamma$
 we substituted the values of the lightest mass (m1/m3)for(NH/IH) as (0.07eV/0.065eV) and best fit values for the parameters ${\Delta m_{sol}}^2$, ${\Delta m_{atm}}^2$,
 $\rm\delta$, $\rm\theta_{13}$, while varying both the Majorana phases, $\alpha, \beta$ from 0 to 2$\pi$ and thus 
obtained $\rm{ M_\nu}^{II} $ and hence $\rm M_{RR}$ as a function of the atmospheric mixing angle $\theta_{23}$. Thus BR can be obtained as a function of $\sin^2\theta_{23}$
from equation (\ref{eqa}). We have varied the value of $\rm\sin^2\theta_{23}$ in its 3\rm$\sigma$ range as in table \ref{t1} and the lightest neutrino mass from $10^{-3}$ to $10^{-1}$ and obtained the values of BR for different mixing patterns,
TBM, BM, HM, GRM. The variation is shown in figure (\ref{fig12}),(\ref{fig13}),(\ref{fig14}) and (\ref{fig15}) for both NH and IH.

\clearpage

\section{CONCLUSION}{\label{sec:level7}}
The quest for NDBD and its interrelation with neutrino mass makes it a very interesting and enthralling topic of research at present time. Its existence would
not only confirm the intrinsic nature of the neutrinos but also would provide a stringent limit on the absolute scale of the neutrino mass. In this paper, 
we contemplated the implications of NDBD and LFV in LRSM framework. Owing to the presence of new scalars and gauge bosons in this model, various additional sources would 
give rise to contributions to NDBD process, which involves RH neutrinos, RH gauge bosons, scalar Higgs triplets as well as the mixed LH-RH contributions. For a simplified 
analysis we have ignored the left-right gauge boson mixing and heavy light neutrino mixing.  We have considered the extra gauge bosons and scalars to be of the order of TeV.
Again the existence of non zero $\rm \theta_{13}$ has many implications in neutrino sector beyond SM. A simple way to accomodate non zero $\rm \theta_{13}$ is by adding a 
perturbation matrix to the neutrino mass matrix. A well known neutrino mass mixing pattern is the one obeying $\rm \mu-\tau$ symmetry. In our present analysis, we have 
considered the different realizations of the $\mu-\tau$ symmetric mass matrices, namely, TBM, BM, HM, GRM matrices. The perturbation to this matrices to generate non
zero $\theta_{13}$ is obtained from either of the seesaw terms, type I and type II. We have considered two different approaches, 
  type I giving $\rm\mu-\tau$ symmetry and type II as perturbation,  
 type II giving $\rm\mu-\tau$ symmetry and type I as perturbation,
for generation of non zero $\rm\theta_{13}$. 
We analysed the standard as well as new physics contribution to the effective mass $\rm m_{eff}$  governing 
NDBD as well as the half life considering both type I and type II seesaw.  We have shown the variations of the effective mass as well as the half life
with the lightest neutrino mass which corresponds to the standard as well as the non standard contributions. We have seen from our analysis that both the approaches yields
different consequences in NDBD. The various parameters we have chosen for our numerical analysis are consistent with constraints from $\nu$ oscillation experiments. We have
also discussed the impacts of the lightest neutrino mass and not so precisely known atmospheric mixing angle, $\theta_{23}$ on the behaviour of LFV of the decay process, $\rm \mu\rightarrow 3e $ and $\rm\mu\rightarrow e\gamma$ 
respectively.
Based on our observations, the following conclusions could be arrived at,
\begin{itemize}

\item In the standard light neutrino contribution to NDBD, it is observed that  all the mass patterns (TBM, HM, GRM) yields almost similar results for NH
mass spectrum. The effective mass governing NDBD is found to be of the order of $\rm10^{-3}$ eV and  are within and much below the current experimental limit \cite{kamland}.
Whereas in case of IH mass spectrum, for TBM, HM and GRM, the values of effective mass are found to be within and close to the experimental limit and are of the order of $10^{-2}$ eV.
However, in all the cases, the light neutrino contribution can saturate the experimental limit 
for lightest neutrino mass ($\rm m_{1}/m_{3}$) for (NH/IH) of around 0.1 eV.

\item In new physics contribution considering perturbation in type II seesaw, for IH, TBM, HM and GRM shows results within the recent experimental bound for lightest mass
varying from  (0.001-0.1) eV. Whereas, for NH the effective mass lies within experimental limit for lightest mass in the range  (0.01-0.1) eV. In case of half life also, except 
BM mass pattern, TBM, HM and GRM  schemes shows better results. In all the cases, both NH and IH seems to be more compatible with the experimental results.

 \item In new physics contribution considering perturbation in type I seesaw, the values that are consistent with experimental bound imposed by KamLAND-Zen are found for lightest mass
 (0.001-0.1) eV for TBM and about (0.01-0.1 eV) for all other cases. Whereas for half life, TBM shows better results. In all other mixing patterns, half life lies within experimental
 bound for values of lightest mass lying from (0.005-0.1) eV for IH. 
\item It is observed from 
our analysis that the  BR for the process $\rm \mu\rightarrow 3e $ in the LRSM remains consistent with the experimental bound for a wide range of light neutrino mass. However,
it depends on the neutrino mass spectrum as evident from fig \ref{fig12} and \ref{fig13}. In case of IH, the BR is spread over a wide range and lies even in the range of the
recently proposed limit with a sensitivity of $10^{-16}$.
 For the process, $\rm \mu\rightarrow e\gamma$, the results for BR are found to be consistent with the experimental limit for all the mixing patterns, except for HM and BM (NH)  in the
 $3\sigma$ range of $\theta_{23}$ . In this case, the dependence of LFV on the neutrino mass spectrum is not much significant as seen in fig \ref{fig14} and \ref{fig15}.
\end{itemize}
\par The effective neutrino mass depends on the character of the neutrino mass spectrum. In most of our analysis in case of NDBD as well as LFV,
we have observed that both the hierachial patterns shows almost equal dominance. However, it is easier to observe the process if we consider the leading order 
mass matrices obeying $\rm \mu$-$\tau$ symmetry
 originating from type I seesaw and using type II seesaw as perturbations to generate non zero $\rm \theta_{13}$. Nevertheless, a more detailed analysis considering the presence of all the mechanisms
which can generate the process in the LRSM framework should be persued to give a general conclusion.
\section{APPENDIX}{\label{sec:level8}}
\textbf{Elements of the type II Seesaw mass matrix (case B) and type I Seesaw mass matrix (Case C):}
\begin{equation}
 S_{11}=\left(c^2_{12}c^2_{13}-X{c_{12}^2}^{\mu\tau}\right)m_1+e^{2i(\beta-\delta)}s^2_{13}m_3+\left(c^2_{13}s^2_{12}-X{s_{12}^2}^{\mu\tau}\right)e^{2i\alpha}m_2
\end{equation}
\begin{equation}
\begin{split}
S_{12}=\left(-c_{12}c_{13}c_{23}s_{12}-c^2_{12}c_{13}s_{13}s_{23}e^{i\delta}+X{c_{12}^{\mu\tau}}{c_{23}^{\mu\tau}}{s_{12}^{\mu\tau}}\right)m_1+\\
\left(-c_{13}s_{12}c_{12}c_{23}e^{2i\alpha}-c_{13}s^2_{12}s_{13}s_{23}e^{i(2\alpha+\delta)}+X{c_{12}^{\mu\tau}}{c_{23}^{\mu\tau}}{s_{12}^{\mu\tau}}e^{2i\alpha}\right)m_2+\\
\left(c_{13}s_{13}s_{23}e^{i(2\beta-\delta)}\right)m_3
\end{split}
\end{equation}
\begin{equation}
\begin{split}
 S_{13}=\left(c^2_{12}c_{13}c_{23}s_{13}e^{i\delta}+s_{12}s_{23}c_{12}c_{13}-X{c_{12}^{\mu\tau}}{s_{12}^{\mu\tau}}{s_{23}^{\mu\tau}}\right)m_1+\\
 \left(-c_{13}s_{12}c_{23}s_{12}s_{13}e^{i(2\alpha+\delta)}-X{c_{12}^{\mu\tau}}{s_{12}^{\mu\tau}}{s_{23}^{\mu\tau}}e^{2i\alpha}\right)m_2+\\
 \left(e^{i(2\beta-\delta)}c_{13}c_{23}s_{13}\right)m_3
 \end{split}
\end{equation}
\begin{equation}
\begin{split}
 S_{21}=\left(-c_{12}c_{13}c_{23}s_{12}-c^2_{12}c_{13}s_{13}s_{23}e^{i\delta}+X{c_{12}^{\mu\tau}}{c_{23}^{\mu\tau}}{s_{12}^{\mu\tau}}\right)m_1+\\
 \left(c_{13}s_{12}c_{12}c_{23}e^{2i\alpha}-s^2_{12}s_{13}s_{23}c_{13}e^{i(2\alpha+\delta)}+X{c_{12}^{\mu\tau}}{c_{23}^{\mu\tau}}{s_{12}^{\mu\tau}}e^{2i\alpha}\right)m_2\\
 \left(e^{i(2\beta-\delta)}c_{13}s_{23}s_{13}\right)m_3
 \end{split}
\end{equation}
\begin{equation}
 \begin{split}
  S_{22}=\left({\left(c_{23}s_{12}-e^{i\delta}c_{12}s_{13}s_{23}\right)}^2-X{c_{23}^2}^{\mu\tau}{s_{12}^2}^{\mu\tau}\right)m_1+\\
  \left(-X{c_{12}^2}^{\mu\tau}{c_{23}^2}^{\mu\tau}+{\left(-c_{12}c_{23}-e^{i\delta}s_{12}s_{13}s_{23}\right)}^2\right)m_2e^{2i\alpha}+\\
  \left(c^2_{13}s^2_{23}-X{s_{23}^2}^{\mu\tau}e^{2i\beta}\right)m_3
  \end{split}
\end{equation}
\begin{equation}
 \begin{split}
S_{23}= \left(\left(-c_{12}c_{23}s_{13}e^{i\delta}+s_{12}s_{23}\right)\left(-c_{23}s_{12}-e^{i\delta}c_{12}s_{13}s_{23}\right)+X{c_{23}^{\mu\tau}}{s_{12}^2}^{\mu\tau}{s_{23}^2}^{\mu\tau}\right)m_1+\\
 \left(\left(-e^{i\delta}c_{23}s_{12}s_{13}+c_{12}s_{23}\right)\left(-c_{12}c_{23}-e^{i\delta}s_{12}s_{13}s_{23}\right)+X{c_{12}^2}^{\mu\tau}{c_{23}^{\mu\tau}}{s_{23}^{\mu\tau}}\right)m_2e^{2i\alpha}+\\
 \left(c^2_{13}c_{23}s_{23}e^{2i\beta}-{c_{23}^{\mu\tau}}{s_{23}^{\mu\tau}}\right)m_3
 \end{split}
\end{equation}
\begin{equation}
 \begin{split}
S_{31}=\left(c^2_{12}c_{13}c_{23}s_{13}e^{i\delta}+s_{12}s_{23}c_{12}c_{13}-X{c_{12}^{\mu\tau}}{s_{12}^{\mu\tau}}{s_{23}^{\mu\tau}}\right)m_1+\\
  \left(c_{13}s^2_{12}e^{i\delta}c_{23}s_{13}+c_{12}s_{23}c_{13}s_{12}e^{2i\alpha}-X{c_{12}^{\mu\tau}}{s_{12}^{\mu\tau}}{s_{23}^{\mu\tau}}\right)m_2e^{2i\alpha}+\\
  \left(e^{2i\beta-i\delta}c_{13}c_{23}s_{13}\right)m_3
 \end{split}
\end{equation}
\begin{equation}
 \begin{split}
S_{32}=\left(\left(-e^{i\delta}c_{12}c_{23}s_{13}+s_{12}s_{23}\right)\left(-c_{23}s_{12}-e^{i\delta}c_{12}s_{13}s_{23}\right)+{c_{23}^{\mu\tau}}{s_{12}^2}^{\mu\tau}{s_{23}^{\mu\tau}}\right)m_1\\
 \left(\left(-e^{i\delta}c_{23}s_{12}s_{13}+c_{12}s_{23}\right)\left(-c_{12}c_{23}-e^{i\delta}s_{12}s_{13}s_{23}\right)+X{c_{12}^2}^{\mu\tau}{c_{23}^{\mu\tau}}{s_{23}^{\mu\tau}}\right)e^{2i\alpha}m_2\\
  \left(c^2_{13}c_{23}s_{23}-X{c^{\mu\tau}}_{23}{s^{\mu\tau}}_{23}\right)e^{2i\beta}m_3
 \end{split} 
\end{equation}
\begin{equation}
 \begin{split}
S_{33}=\left({\left(-e^{i\delta}c_{12}c_{23}s_{13}+s_{12}s_{23}\right)}^2-X{s_{12}^2}^{\mu\tau}{s_{23}^2}^{\mu\tau}\right)m_1+\\
  \left({\left(-e^{i\delta}c_{23}s_{12}s_{13}+c_{12}s_{23}\right)}^2-X{c_{12}^2}^{\mu\tau}{s_{23}^2}^{\mu\tau}\right)e^{2i\alpha}m_2+\\
  \left(c^2_{13}c^2_{23}-{c_{23}^2}^{\mu\tau}\right)e^{2i\beta}m_3
 \end{split}
\end{equation}
Where, $\rm {c_{ij}^{\mu\tau}}= \cos\theta_{ij}^{\mu\tau}$, $\rm {s_{ij}^{\mu\tau}}=\sin\theta_{ij}^{\mu\tau}$ represents the mixing angles for $\rm \mu-\tau$ symmetric neutrino mass matrix (TBM, BM, HM, GRM).


\begin{thebibliography}{99}


\bibitem{minos}J. J. Evans, [arXiv:1307.0721 [hep-ex]] (2013).

\bibitem{t2k}K. Abe et al. [T2K Collaboration], Phys. Rev. Lett. \textbf{107}, 041801 (2011), [arXiv:1106.2822[hep-ex]].

\bibitem{doublechooz}Y. Abe et al., Phys. Rev. Lett. \textbf{108}, 131801 (2012).

\bibitem{dayabay}F. P. An et al., [DAYA-BAY Collaboration], Phys. Rev. Lett. \textbf{108}, 171803 (2012).

\bibitem{reno}J. K. Ahn et al., [RENO Collaboration], Phys. Rev. Lett. \textbf{108}, 191802 (2012).

\bibitem{sigma}D. V. Forero, M. Tortola and J. W. F. Valle, Phys. Rev.  \textbf{D90}, 093006(2014).
 
\bibitem{planck}P. A. R. Ade et al., [PLANCK collaboration], Astron.Astrophys. \textbf{571}, A16 (2014).

\bibitem{planck1}P. A. R.Ade et al., [Planck Collaboration], Astron.Astrophys. 594 \textbf{A13}, [arXiv:1502.01589].

\bibitem{type1}P. Minkowski, Phys. Lett. \textbf{B67}, 421 (1977); M. Gell-Mann, P. Ramond, and R. Slansky (1980), print-80-0576 (CERN); T. Yanagida (1979), in 
Proceedings of theWorkshop on the Baryon Number of the Universe and Unified Theories, Tsukuba, Japan, 13-14 Feb 1979; R.N. Mohapatra and G. Senjanovic, Phys. Rev. Lett  \textbf{44},
912 (1980).

\bibitem{type2}R. N. Mohapatra and G. Senjanovic, Phys. Rev. \textbf{D23}, 165 (1981); R. N. Mohapatra, Nucl. Phys. Proc. suppl. \textbf{138}, 257 (2005); S. Antusch and S. F. King,
Phys. Lett. \textbf{B597}, (2), 199 (2004); Brahmachari and R. N. Mohapatra, Phys. Rev. \textbf{D58}, 015001 (1998).

\bibitem{type3}R. Foot, H. Lew, X. G. He and G. C. Joshi, Z. Phys. \textbf{C44}, 441 (1989).
\bibitem{inverse} Mohapatra, R.N. Mechanism for understanding small neutrino mass in superstring theories  Phys. Rev. Lett. \textbf{56}, 561-563, (1986).
\bibitem{LRSM}J. C. Pati and A. Salam, Phys. Rev. \textbf{D10}, 275 (1974); R. N. Mohapatra and J. C. Pati, Phys. Rev. \textbf{D11}, 2558 (1975); G. Senjanovic and R. N. Mohapatra, 
Phys. Rev. \textbf{D12}, 1502 (1975);  R.N. Mohapatra and R. E. Marshak, Phys. Rev. Lett. \textbf{44}, 1316 (1980); G. Senjanovic, Nucl. Phys. \textbf{B153} 334-364, (1979) .

\bibitem{NDBD} W. Rodejohann, Int. J. Mod. Phys.\textbf{ E20} (2011); S. M. Bilenky, 
C. Giunti, Mod. Phys. Lett. \textbf{A27}, 1230015, (2012).
\bibitem{maj}J. Schechter and J. W. F. Valle, Phys. Rev. \textbf{D25}, 2951 (1982).
\bibitem{table}Bernhard Schwingenheuer, [arXiv:1201.4916v1 [hep-ex]] (2012).
\bibitem{alpha}S.M. Bilenky, J. Hosek and S.T. Petcov, Phys. Lett. \textbf{B94}, 495 (1980) ; M. Doi et al., Phys. Lett. \textbf{B102}, 323 (1981) ; C. Giunti,
Phys. Lett. \textbf{B686 }, 41 (2010), [arXiv:1001.0760]; R. Samanta, M. Chakraborty, A. Ghoshal, [arXiv:1502.06508 [hep-ph]] (2016).
\bibitem{P}  R. N. Mohapatra and W. Rodejohann, Phys. Lett. \textbf{B644},59 (2007).

\bibitem{ndbd}A. Garfagnin, Int. J. Mod. Phys. Conf. Ser., 31, 1460286, (2014); Igor Ostrovskiy, Kevin O'Sullivan, Mod. Phys. Lett.\textbf{A31}, 1630017 (2016).

\bibitem{kamland}A. Gando et. al., [KamLAND-Zen Collaboration], Phys. Rev. Lett. \textbf{117}, 082503 (2016).
\bibitem{gerda}M. Agostini et. al., [GERDA Collaboration], Nuclear and Particle Physics Proceedings (273-275), 1876-1882, (2016).
\bibitem{mutau} H. Ishimori, T. Kobayashi, H. Ohki, Y. Shimizu, H. Okada and M. Tanimoto, Prog. Theor. Phys. Suppl. 183, 1 (2010); S.F. King and C. Luhn, Rept. Prog. Phys. 76,
 056201; G. Altarelli and F. Feruglio, Nucl. Phys. \textbf{B741}, 215 (2006) [hep-ph/0512103]; E. Ma and D. Wegman, Phys. Rev. Lett. \textbf{107}, 061803 (2011); P.S. Bhupal Dev,
B. Dutta, R. N. Mohapatra and M. Severson, Phys. Rev. D \textbf{86} 035002 (2012); G. Altarelli, F. Feruglio, Nucl. Phys. \textbf{B 720}, 64 (2005); G. Altarelli, F. Feruglio,
Nucl. Phys. \textbf{B 720}, 64 (2005); G. Altarelli, F. Feruglio, L. Merlo and E. Stamou, JHEP \textbf{08}, 021 (2012);
M. Borah, D.Borah, M.K Das and S.Patra, Phys. Rev. \textbf{D90}, 095020 (2014), [arXiv:1408.3191]; F. Vissani, hep-ph/9708483, (1997); V. D. Barger, S. Pakvasa, 
T. J. Weiler and K. Whisnant, Phys. Lett. \textbf{B437} (1), 107, (1998); P. F. Harrison, D. H. Perkins and W. G. Scott, Phys. Lett. \textbf{B530}, 167 (2002); P. F. Harrison and W. G. Scott, Phys. Lett.\textbf{B535},
163 (2002); Z. Z. Xing, Phys. Lett.\textbf{B533}, 85 (2002); L. L. Everett and A. J. Stuart, Phys. Rev. \textbf{D79}, 085005 (2009); T.Fukuyama, H.Nishiura, [arXiv:hep-ph/9702253]; T. Fukuyama, PTEP 2017, 3, 033B11 (2017);
J.Schechter, J.W.F. Valle, Phys. Rev. \textbf{D22}, 2227 (1986); J.Schechter, J.W.F. Valle, Phys. Rev. \textbf{D25}, 774 (1982); D.C. Rivera-Agudelo, A Perez-Lorenzana, 
Phys.Rev. \textbf{D92} 7, 073009 (2015);  D.C. Rivera-Agudelo, A Perez-Lorenzana, Phys. Lett. \textbf{B760} 153 (2016).

\bibitem{nonzero1} D. Borah, Nucl. Phys.\textbf{B876} (2), 575-586, (2013); 
D. Borah, S. Patra, P. Pritimita, Nucl. Phys. \textbf{B881}, 444-446, (2014); J.C Gomez-Izquierdo; [arXiv: 1701.01747]
 
\bibitem{lfv1}  V. Cirigliano, A. Kurylov, M. J. Ramsey-Musolf and P. Vogel, Phys. Rev. \textbf{D70}, 075007, (2004); J. Barry and W. Rodejohann, JHEP \textbf{1309}, 153 (2013).
\bibitem{lfv5}D. Borah  and A. Dasgupta, JHEP \textbf{07}, 022 (2016).


\bibitem{ls1}
R. L. Awasthi, P. S. Bhupal Dev and M. Mitra, Phys. Rev  \textbf{D93}, 011701 (2016);  P.Gu, JHEP \textbf{09} 152, (2016);
G. Bambhaniya, P. S. Bhupal Dev, S. Goswami, M. Mitra, JHEP \textbf{04} 046, (2016); P.S Bhupal Dev, S.Goswami, M.Mitra and W. Rodejohann, Phys. Rev. \textbf{D88}, 091301 (2013);
M. K. Parida, S. Patra, Phys. Lett. \textbf{B718} (2013); S. Ge, M. Lindner and S. Patra
, JHEP \textbf{1510}, 077, (2015), [arXiv:1508.07286]; V. Tello et al., Phys. Rev, Lett. \textbf{1069}151801 (2011) ; P.vPritimita, N. Dash, S. Patra, [arXiv:1607.07655][hep-ph]; M.Lindler, F.S. Queiroz et al.,
JHEP \textbf{06} (2016)140; S.F Ge, M.Lindler and S.Patra, JHEP \textbf{1510},  077 (2015), [arXiv:1508.07286].
\bibitem{perturbations} M. Lindler, W. Rodejohann, JHEP\textbf{0705},089 (2007); W. Rodejohann, Phys.Rev.\textbf{D70}, 073010 (2004).
\bibitem{ndbddb2}J. Chakraborty, H.Z.Devi, S.Goswami, S.Patra, JHEP  \textbf{1208}, 008 (2012).
\bibitem{ndbddb1}D. Borah, A. Dasgupta, JHEP \textbf{1511}, 208 (2015).
\bibitem{higgs} R. N. Mohapatra and J.D. Vergados, Phys. Rev. Lett.\textbf{47}, 1713 (1981); C.E.Picciotto and M.S. Zahir,Phys. Rev.\textbf{D26}, 2320 (1981).
\bibitem{electroweak}A. Pich, [arXiv:hep-ph/0502010] (2005).
\bibitem{genericlrsm}R. N. Mohapatra, 3rd edition (Springer, New York,2003).
\bibitem{breaking}D. Borah, Phys. Rev. \textbf{D83}, 035007 (2011).
\bibitem{dbd1} M. Hirsch, H.V. Klapdor-Kleingrothaus, O. Panella, Phys. Lett. \textbf{B374},(1996).
\bibitem{nonzero}Asan Danamik, Journal of Phys:Conference Series 539,  012012 (2014).
\bibitem{momentum}R. L. Awasthi, P.S. Bhupal Deb, M. Mitra, Phys. Rev. \textbf{D93} (1), 011701 (2016), [arXiv:1509.05387[hep-ph]].
\bibitem{muegamma}A. M. Baldini et al., [MEG collaboration], Eur. Phys. J. \textbf{C76} (8), 434 (2016).
\bibitem{SINDRUM}U. Bellgardt et al., [SINDRUM collaboration], Nucl.Phys.\textbf{B299}, 1 (1988).
\bibitem{sindrum2}Andre Schoning on behalf of $\mu 3e$ collaboration  https://indico.cern.ch/event/175067/
call-for-abstracts/102/file/0; A Blonde et al.,[arXiv:1301.6113[physics.ins-det]].
\bibitem{lfv} V. Cirigliano, A. Kurylov, M. J Ramsey-Musof and P. Vogel, Phys. Rev.\textbf{D70}, 075007;
A. Gouvea  and P. Vogel, Nucl. Phys. \textbf{B71}, 75-92(2013) [arXiv:1303.4097[hep-ph]].
\bibitem{GRM} Carl H. Albright, A. Dueck, W.Rodejohann, Eur. Phys. J \textbf{C70}, 1099-1110, (2010), [arXiv:1004.2798].
\bibitem{GRM1} Y. Kajiyama, M. Raidal, A. Strumia, Phys. Rev. \textbf{D76}, 117301 (2007); A. Dutta, F.S. Ling, P. Ramond, Nucl. Phys. \textbf{B671}, 383 (2003).
\bibitem{collider} M. Mitra et al., Phys. Rev \textbf{D94}, 095016 (2016).
\bibitem{newphysics}D. Borah and A. Dasgupta, doi:10.1007/JHEP \textbf{11}, 208 (2015).
\bibitem{phasefactor}J. Kotila and F. Iachello, Phys. Rev. \textbf{C85}, 034316 (2012).
\bibitem{ndbdg}P. S. Bhupal Dev, S. Goswami and M. Mitra, Phys. Rev.\textbf{D91}, 113004 (2015).
\bibitem{nme}G. Pantis, F. Simkovik, J. D. Vergados and A. Faessler, Phys. Rev.\textbf{C53}, 695 (1996).

\end{thebibliography}
\end{document}